\documentclass[twocolumn,amsmath,amssymb,floatfix]{revtex4}

\usepackage{bm}
\usepackage{amsmath}
\usepackage{epsf}

\newcommand{\beq}{\begin{equation}}
\newcommand{\DE}{\rm DE}
\newcommand{\eeq}{\end{equation}}
\newcommand{\beqa}{\begin{eqnarray}}
\newcommand{\eeqa}{\end{eqnarray}}

\newcommand{\etal}{{\it et al. }}

\def\simlt{\lesssim}

\newcommand{\ApJS}{Astrophys. J Supp.}
\newcommand{\ApJL}{Astrophys. J Lett.}
\newcommand{\ApJ}{Astrophys. J}
\newcommand{\PRL}{Phys. Rev. Lett.}
\newcommand{\PRD}{Phys. Rev. D}
\newcommand{\MNRAS}{Mon. Not. R. Astron. Soc.}

\newcommand{\AsAs}{Astron. Astrophys.}

\newcommand{\aut}[2]{{#2.\ #1,}}
\newcommand{\laut}[2]{{#2.\ #1,}}
\newcommand{\refs}[6]{#2, {\bf #3},  {#4} (#5).}
\newcommand{\mrefs}[6]{#2, {\bf #3},  {#4} (#5);}

\newcommand{\mybib}[2]{\bibitem{#2}}

\begin{document}

\title{Self-Consistency and Calibration of\\ Cluster Number Count Surveys for
Dark Energy}
\author{Wayne Hu}
\affiliation{
{}Center for Cosmological Physics, Department of Astronomy and Astrophysics, 
and Enrico Fermi Institute, University of Chicago, Chicago IL 60637
}

\begin{abstract}
\baselineskip 11pt
Cluster number counts offer sensitive probes of the dark energy if and
only if the {\it evolution} of the cluster mass versus observable relation(s) is well calibrated. 
We investigate the potential for internal calibration 
by demanding consistency in the counts as a function of the observable. 
In the context of a constant 
dark energy equation of state, known initial fluctuation amplitude expected 
from the CMB, universal underlying mass function, and an idealized selection, 
we find that the ambiguity from the normalization of 
the mass-observable relationships, or an extrapolation of
external mass-observable determinations from higher masses,  
can be largely eliminated with a sufficiently deep survey, 
even allowing for an arbitrary evolution. 
More generally, number counts as a function of both the redshift and the 
observable enable strong consistency tests
on assumptions made in modelling the mass-observable relations and cosmology. 
\end{abstract}
\maketitle

\section{Introduction}

The number density of massive galaxy clusters is exponentially
sensitive to the amplitude of initially Gaussian density fluctuations and 
has long been recognized as a sensitive cosmological probe 
\cite{Cluster}.  
In particular, 
the evolution of the number counts above a given mass threshold
can determine the properties of the 
dark energy that accelerates the 
expansion \cite{WanSte98,Dieetal01,HaiMohHol01}.  

This potential can only be realized if, in addition to the cluster redshifts,
the cluster masses themselves are known, at least statistically.  
Unfortunately the total mass is not a direct observable and must be 
inferred through scaling relations
with, e.g.\ 
the Sunyaev-Zel'dovich (SZ) flux decrement, 
the $X$-ray flux or temperature, 
weak lensing shear, or optical velocity dispersion.  Indeed the 
normalization of the mass-temperature
relation is currently the leading source of ambiguity in interpreting the local
cluster abundance (e.g.\ \cite{PieScoWhi01,Sel02}).  It compromises dark 
energy constraints if the cluster mass selection cannot be defined to
a few percent in mass \cite{HaiMohHol01}.

The mass-observable relation can potentially be calibrated within a survey itself
if its effect on the number counts is not degenerate with the cosmology.  
For example, with a single cut
on the temperature, the survey itself can calibrate the mass-temperature normalization to
better than the tens of percent that span the determinations in the
current literature, if
it does {\it not} evolve \cite{LevSchWhi02}.  However this method fails
if the relationship has uncertain evolution that 
mimics the cosmology \cite{MajMoh02}.

Because the number density of clusters as a function of mass has
a fixed functional form given by cosmological simulations 
(e.g.\ \cite{Jenetal01,Reeetal03}), cluster number counts as a function of 
the observable in principle have the ability to self-calibrate even
an evolving mass-observable relation.   Here we study the potential
for internal calibration and consistency checks in idealized future cluster
number count surveys. 

\section{Statistical Forecasts}
\label{sec:forecasts}

 The cosmological utility of cluster counts stems from 
the simulation-based
prediction of their comoving differential number density as 
 a function of mass \cite{Jenetal01},
\begin{equation}
{d \bar n \over d\ln M} = 0.3 {\rho_{m} \over M} {d \ln \sigma^{-1} \over d\ln M}
        \exp[-|\ln \sigma^{-1} + 0.64|^{3.82}]\,.
\label{eqn:massfun}
\end{equation}
Here $\sigma^2(M,z)$ is the
variance in the linear density field smoothed with a top hat that encloses 
$M$ at the mean matter density today $\rho_{m}$.   The cosmological sensitivity comes
from these quantities and the comoving volume element  in
a redshift slice and solid angle.  

To assess the sensitivity of counts
to various cosmological and mass-observable
scaling parameters $p_{\alpha}$, we employ the Fisher matrix technique 
\cite{HolHaiMoh01,HuKra02}
\begin{equation}
F_{\alpha\beta} = \sum_{ij} 
{\partial \bar n_i \over \partial p_\alpha} 
({\bf C}^{-1})_{ij}
{\partial \bar n_j \over \partial p_\beta} \,,
\end{equation}
where the covariance matrix is given by
\begin{equation}
C_{ij} =\left( \langle n_i n_j \rangle - \bar n_i\bar n_j \right)+ \delta_{ij} \bar n_i/V_i\,.
\end{equation}
The first term represents sample covariance in the volume $V_i$
from large-scale structure, 
calculated as described in \cite{HuKra02}, and the second term shot variance. 
Here 
the number density in a bin 
$n_{i}$ is defined by the redshift 
interval $\Delta z$ around $z_{i}$ and a selection based on some observable 
quantity $f$, such as the SZ flux decrement.
Note that sampling errors for the selections in the same redshift bin
completely covary.

The Fisher matrix is a local approximation to the covariance matrix
of the parameters
\begin{equation}
{\bf C}_{\rm tot} = ({\bf F}+ {\bf C}^{-1}_{\rm prior})^{-1},
\end{equation}
where ${\bf C}_{\rm prior}$ is the covariance matrix from prior information.

The Fisher matrix is evaluated around a fiducial choice of parameters.
For the cosmology, we take: 
the dark energy density $\Omega_{\rm DE}=0.65$, equation of state
$w=-1$, physical matter density $\Omega_m h^2= 0.148$,
physical baryon density $\Omega_b h^2 = 0.02$, tilt $n_s=1$,
and the initial normalization of the 
curvature fluctuations $\delta_\zeta = 4.79 \times 10^{-5}$ at $k=0.01$ 
Mpc$^{-1}$ \cite{Hu01c}
(more conventionally: $\delta_H= 4.42 \times 10^{-5}$, $\sigma_8=0.92$,
or $M_* = 1.2 \times 10^{13} h^{-1} M_\odot$).  
We will however work in a future context where the cosmic microwave background 
(CMB) has
constrained the high redshift universe parameters with 1$\sigma$ errors of
$\sigma(\ln \Omega_{m}h^{2})=\sigma(\ln \Omega_{b}h^{2})=\sigma(n_{s})=
\sigma(\ln \delta_{\zeta})= 0.01$ (e.g.\ \cite{Hu01c}) 
in a flat universe, leaving the dark
energy parameters to be determined by the survey.  

For the mass-observable relation, we take the form 
\begin{equation}
{M_{f} \over M_{0}}  = e^{A(z)} {\left( f \over f_{0} \right)}^{p(z)}\,,
\end{equation}
where $M_{0}$ and $f_{0}$ are dimensional constants and the 
normalization parameter $A(z)$ is dimensionless.  
We take the scaling
parameters as the amplitudes of a piecewise constant form 
$A(z_{i})$ and $p(z_{i})$. 
Specifically, given a variation $\delta A(z_i)$ and $\delta \ln p(z_i)$ from
the fiducial model,
the true
mass is related to the apparent mass as
\begin{equation}
\ln M =  \ln M_{f} + \delta A(z_i) + \delta \ln p(z_i)\, \ln
{ \left( M_{f} \over M_p\right)}\,,
\end{equation}
where $M_p = M_0 e^{A}$ is the pivot mass scale that defines a 
normalization point for possible external information.

This generalizes previous treatments which have assumed constant
normalization \cite{LevSchWhi02} and power law evolution \cite{MajMoh02}.
In the limit of small redshift bins, 
these simpler cases can be recovered from  our more general treatment
by noting that under a re-para\-meter\-iza\-tion of the space
to the set $\pi_\mu(p_\alpha)$, possibly of lower dimension, 
the covariance matrix transforms as
\begin{equation}
C_{\mu\nu} = \sum_{\alpha\beta} {\partial \pi_\mu \over \partial p_\alpha}
                C_{\alpha\beta} {\partial \pi_\nu \over \partial p_\beta}\,.
\label{eqn:reparam}
\end{equation}
The Fisher matrix definition itself can be so viewed.

In the usual approach 
(e.g.\ \cite{HaiMohHol01}), all of the clusters in a given redshift
bin above
a given observable threshold 
are simply binned together.
Clearly, allowing for 
arbitrary variations  $\delta A(z_{i})$
no cosmological information can be extracted.
However the
data at a given redshift may be binned into several {\it apparent}
mass bins based on the observable $f$.  The additional information
supplied by multiple bins allows for a breaking of the
degeneracy.  

Likewise, even if the normalization is fixed by detailed followup 
by $X$-ray or weak lensing 
measurements at some mass $M_p$ (e.g.\ \cite{MajMoh02,HutWhi02}),
an uncertain
scaling index $\delta p(z_{i})$ would again destroy the cosmological 
information with a single bin if $M_p \ne M_{f,\rm min}$.  
Multiple bins again in principle allow the cosmology and the mass-observable relations 
to be jointly determined.

\begin{figure}[tb]
\centerline{\epsfxsize=3.25truein\epsffile{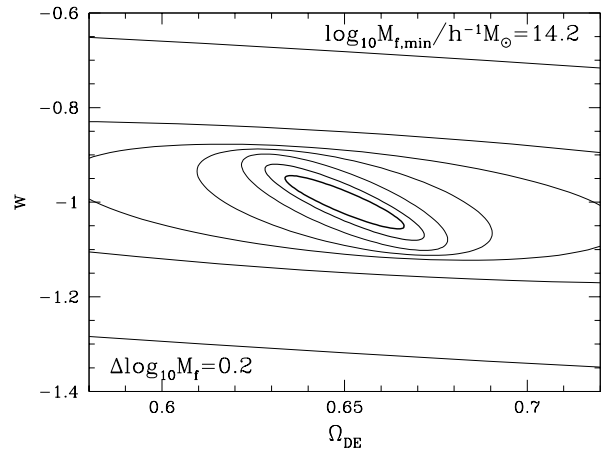}}
\caption{\footnotesize Consistency test with differential counts.  Division of the number counts
into apparent mass bins from a minimum of 
$\log_{10}M_{f,{\rm min}}/h^{-1}{\rm M_\odot}=14.2$ upwards
in steps of $0.2$ (outwards in ellipses) allows several 
nearly independent constraints on the
dark energy that test consistency with the assumed mass-observable 
relation. Strong priors are assumed on the high redshift cosmology
(see text).} 
\label{fig:consistency}
\end{figure}
 
This treatment ignores intrinsic scatter in the mass-observable relation
as well as measurement error in the observable.  More realistically
the selection function is not sharp in mass but assuming its functional
form is well characterized this does not substantially compromise the cosmological 
information \cite{Holetal00,LevSchWhi02}.

For definiteness, let us take a fiducial cluster survey with specifications
similar to the planned South Pole Telescope (SPT) Survey for clusters with
the SZ effect:
an area of 4000
deg$^{2}$ survey and a sensitivity corresponding to a constant  
$\log_{10} M_{f,\rm min}/h^{-1} M_{\odot} = 14.2$ (e.g.\ \cite{MajMoh02}).  
We divide the number counts into bins of redshift $\Delta z=0.1$ out to
$z_{\rm max}=3$ and $\Delta \log_{10} M_{f} = 0.43\Delta\ln M_{f} = 0.2$.  This
crude binning is sufficient to retain the cosmological information.  For
illustration purposes, we take a pivot mass scale of 
$\log_{10} M_p/h^{-1} M_\odot = 14.7$ to reflect potential mass measurement
followup on the high mass end.

\section{Consistency and Calibration}
\label{sec:consistency}

Let us first consider that simulations or mass followup with different observables
have placed strong priors on the mass-observable normalization and
scaling index $\sigma(\delta A(z_{i}))= \sigma(\delta \ln p(z_{i}))= 0$.
In this case, the division into mass bins yields nearly independent
measurements of the dark energy for a consistency check. 

In Fig.~\ref{fig:consistency}, we illustrate the consistency check 
in the $(\Omega_{\DE}, w)$ plane for individual mass bins of $\Delta \log_{10}M_f
=0.2$ from
$\log (M_{f,\rm min}/h^{-1} M_\odot) = 14.2$.
Although the strongest constraints do come from the lowest mass bin,
as expected from the increasing rarity of massive clusters at high redshift, 
the first 6 bins show significant constraints on the dark energy.  
This consistency test demonstrates that there is sufficient information in the
mass binning to attempt some degree of self-calibration of the mass-observable
relations.  

\begin{figure}[tb]
\centerline{\epsfxsize=3.25truein\epsffile{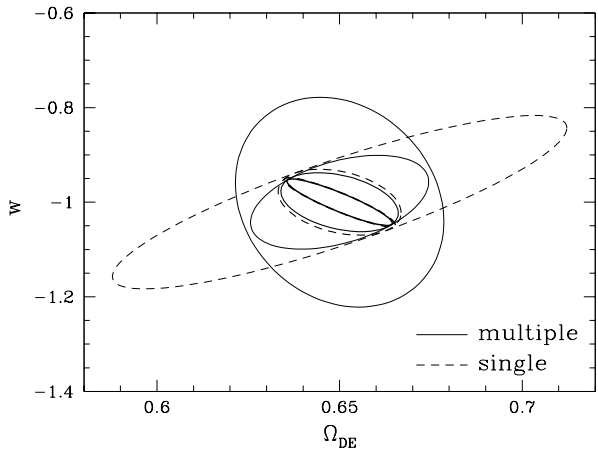}}
\caption{\footnotesize Normalization self-calibration and 
the $68\%$ CL dark
energy constraints.  Solid curves represent binning as in 
Fig.~\ref{fig:consistency},
dashed curves represent a single bin with threshold $M_{f,{\rm min}}$. 
Inner to outer ellipses:
fixed, constant, power law evolution,
arbitrary amplitude $\delta A(z_i)$ (infinite for single mass bin).
Strong priors on the scaling index $p$ are
assumed.}
\label{fig:plotwl}
\end{figure}

Let us next keep the scaling index fixed $p(z_i)$ but allow the
normalization $A(z_i)$ to vary (see Fig. \ref{fig:plotwl}).  Solid
lines represent constraints with the 6 mass bins and dashed lines
those with a single mass bin or mass threshold.  We first allow only
a redshift-independent normalization factor $\delta A(z_i)=\delta A_0$.  
As shown in \cite{LevSchWhi02},
even a single mass bin is sufficient to calibrate the relation to good
enough accuracy for dark energy constraints (here $\sigma(\delta A)=0.05$).
We next employ a power law evolution in the normalization $\delta A(z_i) = \delta A_0 
+ n_a \ln (1+z_i)$.  Here
errors on the dark energy parameters degrade substantially with
a single mass bin \cite{MajMoh02}, but much of the loss is recovered
from the multiple binning.  Finally, we take a fully arbitrary evolution in
the redshift bins, i.e.\ no constraint on the form of $\delta A(z_i)$.  As
expected, there is no constraint on the dark energy with a single mass
bin, but with multiple bins interesting constraints
on the dark energy can still be extracted.  The key here is that
the dark energy evolution is assumed to be smooth and parameterized
by two numbers and so cannot compensate an arbitrary variation in the
normalization evolution.

\begin{figure}[tb]
\centerline{\epsfxsize=3.25truein\epsffile{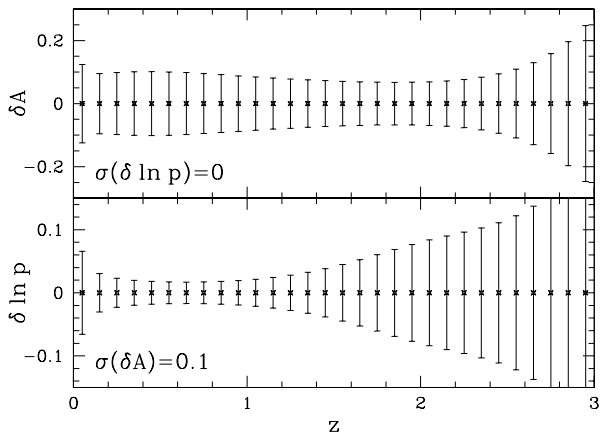}}
\caption{\footnotesize Errors on scaling parameters.  Top: errors on 
$\delta A(z_i)$ assuming fixed scaling index $p$ and the apparent 
mass divisions of Fig.~\ref{fig:plotwl}.  Bottom: errors on 
$\delta \ln p(z_i)$ assuming a prior on the mass-observable 
relation $\sigma(\delta A(z_i))=0.1$ from mass followup at
$\log_{10}M_{p}/h^{-1}{\rm M_\odot}=14.7$
}
\label{fig:Aerror}
\end{figure}

Figure \ref{fig:Aerror} (top) shows the errors on $\delta A(z_i)$ with roughly
10\% constraints on the normalization in each redshift band out
to $z\sim 2$.  Note that the covariance between the bands is nearly
complete so that the overall normalization is still only known
to $\sim 5\%$ but that the evolution is highly constrained. 
Conversely, external information on the mass-observable normalization would
have to be substantially better than $10\%$ to improve on the internal 
results.  In the context of our fiducial cosmology, the critical prior
assumption is that the initial amplitude of the fluctuations is fixed
by the CMB.  Otherwise what is constrained by the multiple mass bins
is a degenerate combination of the initial amplitude and the mass-observable
normalization.  
 
We have so far assumed that the scaling index
$p(z)$ is known a priori but of course that too is part of the
uncertain mass-observable relation.   Allowing for arbitrary variations
$\delta A(z_i)$ {\it and} $\delta \ln p(z_i)$ destroys most of the information
on the dark energy.  However in the less drastic cases of adding a constant
$\delta \ln p$ to a constant $\delta A$ for a two parameter model, 
the additional degradation $\sigma(w)$ is a negligible 
$3\%$ and of the addition of power law variation in $p(z)$, $\delta \ln p = 
\delta \ln p_0  + n_p \ln (1+z_i)$ to
power law variation in $\delta A$ for a 4 parameter model, a factor of $1.76$
for a total of $\sigma(w) = 0.12$.

Another interesting case to consider is if some independent mass calibration, 
say from weak lensing or $X$-ray temperature followup, normalizes the
mass-observable relation on the high mass end. 
Extrapolation down to the survey mass limit can be dangerous due
to the uncertain physics of low mass clusters.
For illustrative
purposes, let us take this constraint as $\sigma(A(z_i)) = 0.1$ 
at the pivot mass of $\log_{10} M_p/h^{-1} M_\odot = 14.7$.  
Despite the constraint on the normalization, if we allow for arbitrary 
variations in the scaling index and retain only a threshold at $>14.2$, no
constraint on the dark energy can be extracted.  With multiple bins,
the errors on $w$ are only degraded by a factor of 1.16
to $\sigma(w)=0.09$ for a completely arbitrary $\delta \ln p(z_i)$.   
Furthermore the errors on $\delta \ln p$ shown in Fig.~\ref{fig:Aerror} (bottom)
are at the several percent level 
and would be of interest in studying cluster physics.

\section{Discussion}
\label{sec:discussion}

As in the case of classical cosmological tests for the dark energy
involving standardized candles and rulers, e.g. supernovae and
the peaks in the CMB power spectrum, cluster number count tests 
require a standardized mass based on observable quantities.  
We have demonstrated that consistency with the well-determined
shape of the mass function from cosmological simulations can in
principle be used to calibrate the survey internally.
Two examples are a normalization that
has an arbitrary evolution in redshift and a scaling from the
more easily calibrated high mass end that has an arbitrary evolution
in the index.  

Our study involves a number of 
idealizations that merit future study and so internal calibration
is best viewed as a useful check on cross-calibration
studies with detailed multi-wavelength followup 
(e.g.\ \cite{MajMoh02,VerHaiSpe02})
and cosmological simulations. 
The crucial assumptions are that the survey be sufficiently deep to
explore a substantial dynamic range in the mass function, 
the underlying mass function is known given a
cosmology,  the 
mass-observable selection function is sharp compared with the binning, the mean
mass-observable relation is power law in form, the high
redshift normalization of the fluctuations is fixed by future CMB 
data, and the dark energy equation of state is constant.

In the fiducial model, a depth comparable to 
that planned for the SPT SZ survey is sufficient to make self-calibration
useful but note that the necessary mass limit scales roughly 
with the non-linear mass scale $M_*$. That depends on the
currently uncertain normalization and so if the normalization is lower
than our fiducial choice it may be beneficial to sacrifice survey
width for depth.    

While the mass function form is based on simulations 
and currently its scaling with cosmology is only known to the
$\sim 10\%$ level \cite{Jenetal01}, determination requires only 
the well-understood gravitational physics of the dark matter 
and is far more secure than simulation-based mass-observable relations.  
In any case, this knowledge is a prerequisite for any cluster number count
study of cosmology.

We have assumed a deterministic mass-observable relation
that makes the selection function sharp in apparent mass.  Gaussian
scatter in this relation does not seriously compromise the cosmological
information \cite{Holetal00,LevSchWhi02} but long uncharacterized
tails to low mass would, given the steep mass function.   Likewise
the cluster physics of preheating and cooling can enter into
the low mass end to make the mass-observable relations deviate from a
power law.  Here the binning serves as a useful consistency test for
the implicitly assumed cluster physics.

The required prior information on the initial normalization is
within reach of the
upcoming CMB satellite missions if the extent of reionization can be
determined (e.g.\ 
\cite{Hu01c,Kapetal02}).  If not,
binning still serves to constrain the evolution of the mass-observable
relation but leaves a degeneracy between the overall normalization 
of the relations and the fluctuations which can be 
fixed, e.g. by CMB lensing,
Ly$\alpha$ forest clustering or even local
cluster abundance studies.  Conversely, external calibration
of the mass-observable relation at the percent level can determine
the fluctuation normalization at a comparable level.  Binning can 
then provide tests of the dark
energy model assumptions \cite{WelBatKne01} or other cosmological 
priors.  It is especially valuable if redshifts are
available only locally (e.g. $z_{\rm max} \simlt 0.6$ from
current optical surveys).

In summary, utilizing the extra information in counting clusters
as a function of both the redshift and their observable properties
allows for joint solutions to the mass-observable relations and cosmology, 
if they are both simple.   If not, it offers valuable 
internal consistency checks against overly simplistic assumptions.

{\it Acknowledgments:} I thank 
Z. Haiman, A.V. Kravtsov, J.J. Mohr, C. Pryke and the Chicago Thunch group for 
useful conversations and J. Valdes for *nix expertise in file recovery.
 WH is supported by NASA NAG5-10840 and 
the DOE OJI program.  

%
%

\end{document}